\documentclass[12pt]{article}

\usepackage{graphicx}   
\usepackage{epsfig}
\begin{document}

\title{Variational Method for the Three-Dimensional Many-Electron Dynamics of Semiconductor Quantum Wells}


\author{Fernando Haas \\ Departamento de F\'{\i}sica \\ Universidade Federal do Paran\'a \\81531-990, Curitiba, PR \\Brazil}
\date{}

\maketitle




\begin{abstract} 
The three-dimensional nonlinear dynamics of an electron gas in a semiconductor quantum well is analyzed in terms of a self-consistent fluid 
formulation and a variational approach. Assuming a time-de\-pen\-dent localized profile for the fluid density and appropriated spatial dependences of the scalar potential and fluid velocity, a set of ordinary differential equations is derived. In the radially symmetric case, the prominent features of the associated breathing mode are characterized. 
\end{abstract}

\maketitle


\section{Introduction}

The fast dynamics of a quantum electron gas confined in nanoscale systems including quantum wells, quantum dots and metal clusters is receiving a great deal of interest \cite{Bloch, Giorgini}. Among possible collective motions, one has the electric dipole response, the so-called Kohn mode \cite{Kohn}, associated to rigid oscillations of the electron gas at the trapping frequency. 

Other fundamental collective oscillations are the breathing modes, characterized by a pulsating electron cloud. Such breathing modes have been investigated through nu\-me\-ri\-cal simulation of the N-body time-dependent Schr\"odinger equation for restricted number of particles, exploring large energy coupling parameter regimes \cite{Bauch}. Soon later, a variational approach based on a mean-field, quantum hydrodynamic model has been used to analytically des\-cri\-be the nonlinear breather dynamics of an one-dimensional electron gas in a semiconductor quantum well \cite{PRB}. A time-dependent Rayleigh-Ritz method was employed in order to reduce the underlying quantum fluid model to a set of nonlinear second-order ordinary differential equations. The numerical analysis of the kinetic, Wigner-Poisson system was shown to be in agreement with the analytical results. In the present work we make a first step toward the extension of this approach to three spatial dimensions. 

\section{Model equations} 

In the mean-field approximation, the electron dynamics can be described by a self-consistent quantum
hydrodynamic model that was also applied for quantum
plasmas \cite{book} and metallic nanostructures
\cite{Crouseilles}. Here only electrostatic fields are considered. Then the evolution of the electron
fluid density $n({\bf r}, t)$ and the electron fluid velocity ${\bf u}({\bf r},t)$ follows from the
continuity and momentum equations, 
\begin{eqnarray}
\label{e1}
\frac{\partial n}{\partial t} &+& \nabla\cdot\,(n{\bf u}) = 0 \,,\\
 \frac{\partial\,{\bf u}}{\partial\,t} &+& {\bf u}\cdot\,\nabla\,{\bf u}  = -\frac{\nabla\,P}{m_* n} 
\label{e2} - \frac{\nabla\,V_{\rm eff}}{m_{*}}
+ \frac{\hbar^2}{2\,m_{*}^2} ~\nabla
\left(\frac{\nabla^2 \sqrt{n}}{\sqrt{n}} \right) ,
\end{eqnarray}
where $m_*$ is the effective electron mass, $\hbar$ Planck's constant divided by $2\pi$, $P = P({\bf r},t)$  the electron fluid pressure and
\begin{equation}
V_{\rm eff} = V_{\rm conf}({\bf r})+V_H({\bf r},t)
\end{equation}
the effective potential, which is composed by a confining $V_{\rm conf}$ and a Hartree $V_H$ terms. The Hartree
potential obeys Poisson's equation, 
\begin{equation}
\nabla^{2}V_H = -
\frac{e^{2}\,n}{\varepsilon} \,,
\end{equation}
where $e$ is the magnitude of the electron
charge and $\varepsilon$ is the effective dielectric per\-mea\-bi\-li\-ty
in the semiconductor quantum well. The term proportional to $\hbar^2$ on the
right-hand side of Eq. (\ref{e2}) is associated to the so-called Bohm potential and is 
responsible for the quantum wave-like effects. 

For the confinement suppose the generally anisotropic potential
\begin{equation}
V_{\rm conf} = \frac{m_* \omega_0^2}{2}\,(\kappa_1 x^2 + \kappa_2 y^2 + \kappa_3 z^2) \,,
\end{equation}
where $\kappa_{1,2,3} > 0$ are dimensionless constants, not necessarily equal.
Moreover, the frequency $\omega_0$
can be related to a fictitious homogeneous positive charge of
density $n_0$ through $\omega_0 = (e^2 n_0/m_* 
\varepsilon)^{1/2} $. 

It is convenient to normalize time to $\omega_0^{-1}$, 
space to $L_{0} = (k_{B}\,T/m_{*})^{1/2}/\omega_{0}$ [where $k_B$ is Boltzmann's constant and $T$ the temperature], density to $n_0$, velocity to
$L_{0}\,\omega_0$, pressure to $n_0\,k_B\,T$ and energy to $k_B\,T$. These are the same rescaled variables as in Refs. \cite{Bauch, PRB}. For simplicity, old and new variables will be represented by the same symbols. 
The model equations then become
\begin{eqnarray}
\label{f1}
\frac{\partial n}{\partial t} + \nabla\cdot\,(n{\bf u}) &=& 0 \,,\\
\label{f2}
 \frac{\partial\,{\bf u}}{\partial\,t} + {\bf u}\cdot\,\nabla\,{\bf u}  &=& -\frac{\nabla\,P}{n} 
- {\bf K}\cdot{\bf r} - \nabla\,V_{H}
+ \frac{H^2}{2} ~\nabla
\left(\frac{\nabla^2 \sqrt{n}}{\sqrt{n}} \right) \,,\\
\label{f3}
\nabla^{2}V_H &=& - n \,,
\end{eqnarray}
where ${\bf K}$ is the diagonal dyad with components $K_{ij} = \delta_{ij}\kappa_j$. 
Quantum effects are measured by the
dimensionless parameter 
\begin{equation}
H = \frac{\hbar\,\omega_{0}}{k_{B}\,T} \,.
\end{equation}
In addition, in the new coordinates we have 
\begin{equation}
V_{\rm conf} = \kappa_1 x^2 + \kappa_2 y^2 + \kappa_3 z^2 \,.
\end{equation}

In order to close the system, the pressure in Eq. (\ref{f2}) must be related to the
electron  fluid density. In the fast time-scale implied by the trapping frequency, an adiabatic equation of state is indicated, since there is no sufficient time for thermalization. In this regard, the temperature parameter $T$ can be interpreted as a measure of the average kinetic energy per electron. Hence we choose the polytropic
relation 
\begin{equation}
P = \overline{n}\,\left(\frac{n}{\overline{n}}\right)^{\gamma} \,,
\end{equation}
where  $\gamma = 5/3$ is the three-dimensional 
polytropic exponent, and $\overline{n}$ a reference 
density, whose choice will be discussed later.  

Let us make an estimate of the parameters in the model. For semiconductor quantum wells \cite{Wijewardane}, typically we have: 
the effective
electron mass and the effective dielectric permeability are,
res\-pec\-ti\-ve\-ly, $m_{*}=0.067m_e$ and $\varepsilon =
13\,\varepsilon_0$ and the equilibrium density is $n_0 = 4.7 \times
10^{22}\,{\rm m}^{-3}$. These values yield an effective plasmon
energy $\hbar\,\omega_0 = 8.62 \,{\rm meV}$, a characteristic
length $L_0 = 16.2 \,{\rm nm}$, a Fermi temperature $T_F=51.8$ K,
and a typical time scale $\omega_0^{-1}=76$ fs. An electron
temperature $T = 200\,{\rm K}$ then corres\-ponds to a value $H =
0.5$.

\section{Time-dependent Rayleigh-Ritz method}

In order to derive a closed system
of ordinary differential equations in which time is the independent variable, 
we first express the quantum hydrodynamical equations in a Lagrangian
formalism. This approach is not based on a
perturbative expansion, and is thus not restricted to the linear
regime. A Lagrangian density corresponding to the system read as 
\begin{eqnarray}
{\cal L} &=& \frac{1}{2}\left(\nabla\,V_H\right)^2 - n\,V_{eff} - \,n\,\frac{\partial\theta}{\partial t} -
\int^{n}W(n')\,dn' \nonumber \\ \label{e4} &-&
\frac{1}{2}\,\left(n\,\left[\nabla\theta\right]^2 + \frac{H^2}{4n}\,\left[\nabla\,n\right]^2\right) \,,
\end{eqnarray}
where the independent fields are $n$, $\theta$, and
$V_H$. The velocity field follows from the auxiliary function
$\theta = \theta({\bf r},t)$ through 
\begin{equation}
{\bf u} = \nabla\theta \,.
\end{equation}
The
quantity $W(n)$ in Eq. (\ref{e4}) originates from the pressure, 
\begin{equation}
W
\equiv \int^{n} \frac{dP}{dn'}\frac{dn'}{n'} =
(5/2)\,(n/\overline{n})^{2/3} \,.
\end{equation}
Taking the variational derivatives of
the action $S = \int\,{\cal L}\,d{\bf r}\,dt$ with respect to $n$,
$\theta$, and $V_H$, we obtain the Eqs. (\ref{f1})--(\ref{f3}).

The existence of a variational formalism can be used to
derive approximate solutions via the time-dependent Rayleigh-Ritz
trial-function method. The electron  density can be taken in the form of an anisotropic Gaussian, 
\begin{equation}
n = \frac{A}{\sigma_1 \sigma_2 \sigma_3}\exp\left(-\frac{R^2}{2}\right) \,,
\end{equation}
while the self-consistent electrostatic field can be taken as
\begin{equation}
V_H = \frac{A\sqrt{\pi/2}}{(\sigma_1 \sigma_2 \sigma_3)^{1/3}}\,\frac{{\rm Erf}\left(R/\sqrt{2}\right)}{R} \,, 
\end{equation}
where the  variable 
\begin{equation}
R = \left[\left(\frac{x-d_1}{\sigma_1}\right)^2 + \left(\frac{y-d_2}{\sigma_2}\right)^2 + \left(\frac{z-d_3}{\sigma_3}\right)^2\right]^{1/2}
\end{equation}
was introduced. Here $\sigma_i$ and $d_i$ are time-dependent functions respectively related to the width and center-of-mass coordinate of the electronic cloud.  In addition, $A$ is a numerical constant directly related to the number of electrons in the quantum well, 
\begin{equation}
A = \frac{1}{(2\pi)^{3/2}} \,\int n \,d{\bf r} \,.
\end{equation}
Finally, Erf is the error function, 
defined by 
\begin{equation}
{\rm Erf}(\mu) = \frac{2}{\sqrt{\pi}}\int_{0}^{\mu}\,e^{-\nu^2}\,d\nu \,.
\end{equation}

The arguments in favour of the selected profile are: it describes a localized solution amenable to relatively simple analytical calculations; the ground state for the three-dimensional anisotropic quantum harmonic oscillator is a Gaussian. Hence we have an exact solution for ne\-gli\-gi\-ble Hartree and pressure terms; in the radially symmetric case ($\sigma_1 = \sigma_2 = \sigma_3$) the proposed form is an exact solution for Poisson's equation. According to the last reasoning, in principle the model becomes less accurate in strongly anisotropic cases. 

To complete the Ansatz, we need to specify the velocity field, which we choose so as to exactly solve the continuity equation for the given $n$. Hence we take the linear form
\begin{equation}
u_i = \frac{\dot\sigma_i}{\sigma_i}(r_i-d_i) + \dot{d}_i \,,\quad i=1,2,3,
\end{equation}
for the $i-$component of the velocity field. 

Since ${\bf u} = \nabla\theta$, the variable $\theta$ in the Lagrangian density can be written as 
\begin{equation}
\theta = \sum_{i=1}^{3}\,\left(\frac{\dot{\sigma}_i}{2\sigma_i}\,(r_i-d_i)^2 + \dot{d}_i\,(r_i-d_i)\right)
\end{equation}
ignoring an irrelevant additive purely time-dependent contribution. 

\section{Mechanical system} 

Now a system of Newton equations for $\sigma_i, d_i$ can be derived. 
Apart from a multiplicative factor, the Lagrangian turns out to be 
\begin{equation}
L = \int{\cal L}\,d{\bf r} = \frac{1}{2}\,\dot{\sigma}^2 + \frac{1}{2}\,\dot{\bf d}^2 - U_{\sigma}({\sigma}) - U_{d}({\bf d}) \,,
\end{equation}
where ${\sigma} = (\sigma_1,\sigma_2,\sigma_3), {\bf d} = (d_1,d_2,d_3)$ and
\begin{eqnarray}
U_\sigma &=& U_{\sigma}({\sigma}) = \frac{1}{2}(\kappa_1 \sigma_{1}^2 + \kappa_2 \sigma_{2}^2 + \kappa_3 \sigma_{3}^2)  \nonumber \\ &+& \frac{H^2}{8}\left(\frac{1}{\sigma_1^2}  + \frac{1}{\sigma_2^2} + \frac{1}{\sigma_3^2}\right) + \frac{\sqrt{2}}{2}\,\frac{A}{(\sigma_1 \sigma_2 \sigma_3)^{1/3}}  \\ &-& \frac{A\,(\sigma_1 \sigma_2 \sigma_3)^{1/3}}{6\sqrt{2}}\,\left(\frac{1}{\sigma_1^2} + \frac{1}{\sigma_2^2} + \frac{1}{\sigma_3^2}\right) + \frac{9}{10}\sqrt{\frac{3}{5}}\left(\frac{A}{\bar{n}}\right)^{2/3}\,\frac{1}{(\sigma_1 \sigma_2 \sigma_3)^{2/3}}  \nonumber \,,\\
U_d &=& U_{d}({\bf d}) = \frac{1}{2}(\kappa_1 d_{1}^2 + \kappa_2 d_{2}^2 + \kappa_3 d_{3}^2)  
\end{eqnarray}
are pseudo-potentials resp. for the breathing and dipole motions. 

Examining $U_\sigma$ we find: the term with $\kappa_{1,2,3}$ is due to the harmonic confinement; the $\sim H^2$ term is due to the Bohm potential; the $~A$ terms are due to the self-consistent Hartree energy; the $\sim (A/\bar{n})^{2/3}$ term is due to the thermodynamic pressure. 

Clearly the dipole and electronic cloud width dynamics are decoupled, a feature of harmonic traps. The dipole motion is linear, 
\begin{equation}
\ddot{d}_i + \kappa_i d_i = 0 \,, \quad i = 1,2,3 \,,
\end{equation}
corresponding to Kohn oscillations. 
On the other hand, the $\sigma_i$ execute coupled nonlinear oscillations.

The equations for the breathing motion read
\begin{eqnarray}
\ddot\sigma_i + \kappa_i\,\sigma_i &=& \frac{\sqrt 2}{6}\,\frac{A}{S\,\sigma_i} - \frac{A\,S}{3\sqrt{2}\,\sigma_{i}^3} + \frac{A}{18\sqrt{2}}\,\frac{S}{\sigma_i}\,\sum_{j=1}^{3}\frac{1}{\sigma_{j}^2} \nonumber \\ 
\label{kepler}
&+& \frac{H^2}{4\,\sigma_{i}^3} + \left(\frac{3}{5}\right)^{3/2}\,\left(\frac{A}{\bar{n}}\right)^{2/3} \frac{1}{S^2 \sigma_i} \,, 
\end{eqnarray}
where 
\begin{eqnarray} 
S &=& S(t) = (\sigma_1 \sigma_2 \sigma_3)^{1/3} \,.
\end{eqnarray}

A more detailed investigation of the dynamical system (\ref{kepler}) will be postponed to future work. Instead, in the following the radial case is considered. 

\section{The isotropic case}

Assuming $\kappa_{1,2,3} = 1, \sigma_{1,2,3} = \sigma$, the breather equation reduces to 
\begin{equation}
\ddot{\sigma} + \sigma = \frac{A}{6\sqrt{2}\sigma^2} + \left(\frac{H^2}{4} + \left(\frac{3}{5}\right)^{3/2}\,\left(\frac{A}{\bar{n}}\right)^{2/3}\right)\,\frac{1}{\sigma^3} \,.
\end{equation}
Before proceeding, we can now define the parameter $\bar{n}$ in the equation of motion, and we do this using thermodynamic arguments. In terms of our Gaussian {\it Ansatz}, we can expect 
\begin{equation}
\bar{n} \sim \frac{A}{\sigma_{0}^3} \,,
\end{equation}
where $\sigma_0$ is the equilibrium value of the variance $\sigma$. Hence $A/\bar{n}$ is not negligible even if $A \to 0$, which means neglecting the self-consistent interaction. Taking the simultaneous limits $A \to 0, H\to 0$ with fixed $A/\bar{n}$ we have  
\begin{equation}
\label{xx}
\ddot{\sigma} + \sigma = \left(\frac{3}{5}\right)^{3/2}\,\left(\frac{A}{\bar{n}}\right)^{2/3}\,\frac{1}{\sigma^3} \,.
\end{equation}

On the other hand, restoring physical coordinates and assuming energy equipartition, we have the ensemble averages 
\begin{equation}
\frac{m_{*}\omega_{0}^2 <x^2>}{2} = \frac{m_{*}\omega_{0}^2 <y^2>}{2} = \frac{m_{*}\omega_{0}^2 <z^2>}{2} = \frac{\kappa_B T}{2} \,,
\end{equation}
which can be also thought as the definition of the temperature parameter $T$. Supposing  
\begin{equation}
<x^2> \quad = \quad <y^2> \quad = \quad <z^2> \quad = \quad \sigma_{0}^2
\end{equation}
we find $\sigma_0 = L_0$. Going back to the rescaled variables, we expect
\begin{equation}
\sigma_0 = 1
\end{equation}
to be the equilibrium value in accordance with energy equipartition, neglecting Hartree and quantum terms. On these grounds, from Eq. (\ref{xx}) we have
\begin{equation}
\bar{n} = \left(\frac{3}{5}\right)^{9/4} \,A \simeq 0.32 A
\end{equation}

In a first approximation, the above choice is adopted also for the $A \neq 0, H \neq 0$ case. Therefore we get
\begin{equation}
\ddot{\sigma} + \sigma = \frac{A}{6\sqrt{2}\sigma^2} + \left(1 + \frac{H^2}{4}\right)\,\frac{1}{\sigma^3} \,.
\end{equation}
or
\begin{equation}
\ddot{\sigma} = -
\frac{dU}{d\sigma} \,,
\end{equation}
where $U(\sigma)$ is a pseudo-potential defined by
\begin{equation}
U =\frac{\sigma^2}{2} + \frac{A}{6\sqrt{2} \sigma} + \frac{1}{2}\,\left(1 + \frac{H^2}{4}\right)\,\frac{1}{\sigma^2} \,.
\end{equation}

\begin{figure}
\includegraphics[height=.3\textheight]{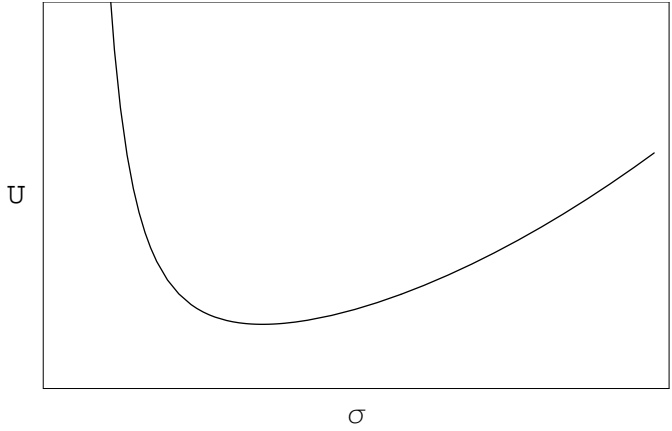}
\caption{Qualitative shape of the pseudo-potential $U(\sigma)$ for arbitrary $A > 0, H > 0$} 
\label{fig1}
\end{figure}

From the shape of the pseudo-potential (see Fig.
\ref{fig1}), it follows that $\sigma$ will always execute
nonlinear oscillations around the unique minimum 
$\sigma_0 = \sigma_0(A,H)$, which is a solution of the algebraic equation
$U'(\sigma_0)=0$, or
\begin{equation}
\label{s}
1 - \frac{A}{6\sqrt{2}\sigma_{0}^3} - \left(1 + \frac{H^2}{4}\right)\frac{1}{\sigma_{0}^4} = 0 \,.
\end{equation}

The frequency
$\Omega=\Omega(A,H)$ of the breather mode, corresponding to the
oscillations of $\sigma$, can be obtained by linearizing the
equation of motion in the vicinity of the 
minimum of $U(\sigma)$. In other words, we use $\sigma_0$ from Eq. (\ref{s}) plugging it in 
\begin{eqnarray}
\label{ss}
\Omega^2 = \left(\frac{d^2 U}{d\sigma^2}\right)_{\sigma=\sigma_0} &=& 1 + \frac{A}{3\sqrt{2}\sigma_{0}^3} + 3 \left(1 + \frac{H^2}{4}\right)\frac{1}{\sigma_{0}^4} \\ &=&    4 - \frac{A}{6\sqrt{2}\sigma_{0}^3} \,, \nonumber
\end{eqnarray}
the last equality following from Eq. (\ref{s}) and showing that $\Omega \rightarrow 2$ in the absence of self-consistent interaction. 

A consistency check can be performed neglecting the Hartree potential, which yields the equilibrium variance 
\begin{equation}
\sigma_0 = \left(1+\frac{H^2}{4}\right)^{1/4} \,. 
\end{equation}
This expression displays the correct low- and high-temperature limits for the
quantum harmonic oscillator: $\sigma_0 \to 1$ for $H \to 0$; and $\sigma_0 \sim \sqrt{H/2}$ for $H >> 1$. These estimates can be found from standard calculations on the canonical ensemble of quantum harmonic oscillators \cite{Imre}. 

Since from Eq. (\ref{ss}) we automatically have $\Omega^2 > 0$ we conclude that 
\begin{equation}
\sigma_{0}^3 > \frac{A}{8\sqrt{2}} \,,
\end{equation}
which further confirms the repulsive effect of the Hartree potential. 

On the other hand,  Eq. (\ref{s}) shows that for large Hartree parameter we have 
\begin{equation}
\sigma_{0}^3 \sim \frac{A}{6\sqrt{2}} \,, 
\end{equation}
Using this in Eq. (\ref{ss}) we find that $\Omega \sim \sqrt{3}$ for large $A$. 

%
%

We can define
\begin{equation}
<n> = \frac{\int n^2 d{\bf r}}{\int n d{\bf r}} = \frac{1}{2\sqrt{2}}\,\frac{A}{\sigma_{0}^3} 
\end{equation}
as the average electron fluid density, where the average is calculated using the density iteself, at the equilibrium value $\sigma = \sigma_0$. Using $n$ for calculating expectation values is reasonable since it corresponds to the quantum probability density function. 

Table \ref{table1} shows that the breather frequency depends
weakly on the parameter $H$ (and hence on the electron fluid 
temperature). The density $<n>$ is shown to decrease with $H$. This follows since quantum diffraction effects tend to enlarge the width of the electronic cloud, which for fixed number of electrons means a smaller density. 

\begin{center}
%
\begin{table}[htbp]
\begin{tabular}{lccc}
{$H$} & {$\sigma_0$} & {$\Omega$} & {$<n>$} \\ \hline
0.0 & 1.03  & 1.97  & 0.32 \\
0.5 & 1.04  & 1.97  & 0.31 \\
1.0 & 1.08  & 1.98  & 0.28 \\
1.5 & 1.14  & 1.98  & 0.24 \\
2.0 & 1.21  & 1.98  & 0.20 \\
2.5 & 1.28  & 1.99  & 0.17 \\
3.0 & 1.36  & 1.99 & 0.14 \\ \hline
\end{tabular}
\caption{Equilibrium width $\sigma_0$, breather frequency $\Omega$ and mean particle density $<n>$ for $A=1$ and various values of $H$}
\label{table1}
\end{table}
\end{center}

Table \ref{table2} shows the dependence of the breather frequency on the Hartree parameter $A$, for fixed $H = 0.5$ which is representative of realistic semiconductor quantum wells. In addition, the average particle density is shown.  

%
\begin{table}[htbp]
\begin{tabular}{lccc}
{$A$} & {$\sigma_0$} & {$\Omega$} & {$<n>$} \\ \hline
0.0 &  1.02  & 2.00  & 0.00 \\
1.0 &  1.04  & 1.97  & 0.31 \\
2.0 &  1.07  & 1.95  & 0.58 \\
3.0 &  1.10  & 1.93  & 0.80 \\
4.0 &  1.12  & 1.92  & 1.00 \\ \hline
\end{tabular}
\caption{Equilibrium width $\sigma_0$, breather frequency $\Omega$ and mean particle density $<n>$ for $H = 0.5$ and various values of $A$}
\label{table2}
\end{table}

\section{Conclusion}

In this work, we derived the basic equations for the three-dimensional variational des\-crip\-tion for the many-electron of a quantum electron gas in a semiconductor quantum well. The first results on the associated linear breathing frequencies were then obtained. The analytic calculations should be confronted against numerical simulation of the Wigner-Poisson system, as made in the one spatial dimension case in \cite{PRB}. In particular, the adequacy of the Gaussian profile proposed here in the case of large Hartree energy should be investigated. Moreover, the effect of the coupled dipole-breather dynamics in non-parabolic wells is also an interesting issue, since in this case the breather mode can be triggered using a purely dipolar
excitation \cite{PRB}. Standard pump-probe experiments can then be used to optically
detect the breather mode.


\vskip.3cm

{\bf Acknowledgments}
\vskip.3cm  
This work was supported by CNPq (Conselho Nacional de Desenvolvimento Cient\'{\i}fico e Tecnol\'ogico). The author also thanks Professor Giovanni Manfredi for useful discussions.

\end{document}